%% file: main.tex
\documentclass[runningheads]{llncs}
\usepackage{graphicx}
\usepackage{booktabs}
\usepackage{numprint}

\usepackage{ulem} 

\usepackage{appendix}

\usepackage[shortlabels]{enumitem}

\usepackage{nameref}

\usepackage{url}

\usepackage{pifont}
\newcommand{\cmark}{\ding{51}}

\usepackage[misc]{ifsym}

\usepackage{acronym}
\newacro{cots}[COTS]{Commercial Off The Shelf}   
\newacro{iiot}[IIoT]{Industrial Internet of Things}   
\newacro{iot}[IoT]{Internet of Things}     
\newacro{it}[IT]{Information Technology}
\newacro{ot}[OT]{Operation Technology}
\newacro{ip}[IP]{Internet Procotol}  
\newacro{hmi}[HMI]{Human Machine Interface}
\newacro{plc}[PLC]{Programmable Logic Controller}  
\newacro{cnc}[C\&C]{Command \& Control}  
\newacro{ics}[ICS]{Industrial Control System} 
\newacro{ids}[IDS]{Intrusion Detection System}
\newacro{scada}[SCADA]{Supervisory Control And Data Acquisition}  
\newacro{swat}[\textit{SWaT}]{\textit{Secure Water Treatment}}  
\newacro{lstm}[LSTM]{Long Short-Term Memory}
\newacro{cps}[CPS]{Cyber-Physical System}
\newacro{wsn}[WSN]{Wireless Sensor Networks}
\newacro{arima}[ARIMA]{Auto-Regressive Integrated Moving Average}
\newacro{dos}[DoS]{Denial of Service}    
\newacro{uv}[UV]{Ultraviolet}    
\newacro{ro}[RO]{Reverse Osmosis}    
\newacro{svm}[SVM]{Support Vector Machine}    
\newacro{sssp}[SSSP]{Single Stage Single Point}
\newacro{ssmp}[SSMP]{Single Stage Multi Point}
\newacro{mssp}[MSSP]{Multi Stage Single Point}
\newacro{msmp}[MSMP]{Multi Stage Multi Point}
\newacro{mtu}[MTU]{Master Terminal Unit}
\newacro{cf}[CF]{Continuous Feedback}
\newacro{ci}[CI]{Continuous Integration}
\newacro{cd}[CD]{Continuous Deployment}
\newacro{kpi}[KPI]{Key Performance Indicator}
\newacro{waf}[WAF]{Web Application Firewall}
\newacro{mpus}[MPU's]{microprocessor units}
\newacro{mcus}[MCU's]{microcontroller units}
\newacro{se}[SE]{secure element}
\newacro{pki}[PKI]{public key infrastructure}
\newacro{sme}[SME]{small and medium-sized enterprise}
\newacro{scratch}[SCRATCh]{SeCuRe and Agile Connected Things}
\newacro{enisa}[ENISA]{European Union Agency for Cybersecurity}

\begin{document}
\title{Creating it from SCRATCh:\\A Practical Approach for Enhancing the Security of IoT-Systems in a DevOps-enabled Software Development Environment}
%
%
\author{Simon D Duque Anton\inst{1}${\textrm{\Letter}}$, Daniel Fraunholz\inst{1}, Daniel Krohmer\inst{1}, Daniel Reti\inst{1}, Hans D Schotten\inst{1}, Franklin Selgert \inst{2}, Marcell Marosv\"olgyi\inst{2},  Morten Larsen\inst{2}, Krishna Sudhakar\inst{3}, Tobias Koch \inst{3}, Till Witt\inst{3}, Cédric Bassem \inst{4}}
\authorrunning{Duque Anton et al.}
\titlerunning{Creating it from SCRATCh}
%
\institute{German Research Center for Artificial Intelligence  (DFKI), Germany \\
\email{\{firstname\}.\{lastname\}@dfki.de} \and
AnyWi, Netherlands \\
\email{\{firstname\}.\{lastname\}@anywi.com} \and
consider it GmbH, Germany \\
\email{\{lastname\}@consider-it.de} \and
NVISO, Belgium \\
\email {cbassem@nviso.eu}
}
\maketitle              
\begin{abstract}
DevOps describes a method to reorganize the way different disciplines in software engineering work together to speed up software delivery.
However,
the introduction of DevOps-methods to organisations is a complex task.
A successful introduction results in a set of structured process descriptions.
Despite the structure,
this process leaves margin for error:
Especially security issues are addressed in individual stages,
without consideration of the interdependence.
Furthermore,
applying DevOps-methods to distributed entities,
such as the Internet of Things (IoT) is difficult as the architecture is tailormade for desktop and cloud resources.
In this work,
an overview of tooling employed in the stages of DevOps processes is introduced.
Gaps in terms of security or applicability to the IoT are derived.
Based on these gaps,
solutions that are being developed in the course of the research project SCRATCh are presented and discussed in terms of benefit to DevOps-environments.


\end{abstract}

\input{body}

%
%
%
\bibliographystyle{splncs04}
\bibliography{bibliography}
\newpage
\appendix
\section{Appendix}
\begin{figure}[ht]
\centering
\includegraphics[angle=90, height=0.715\textheight]{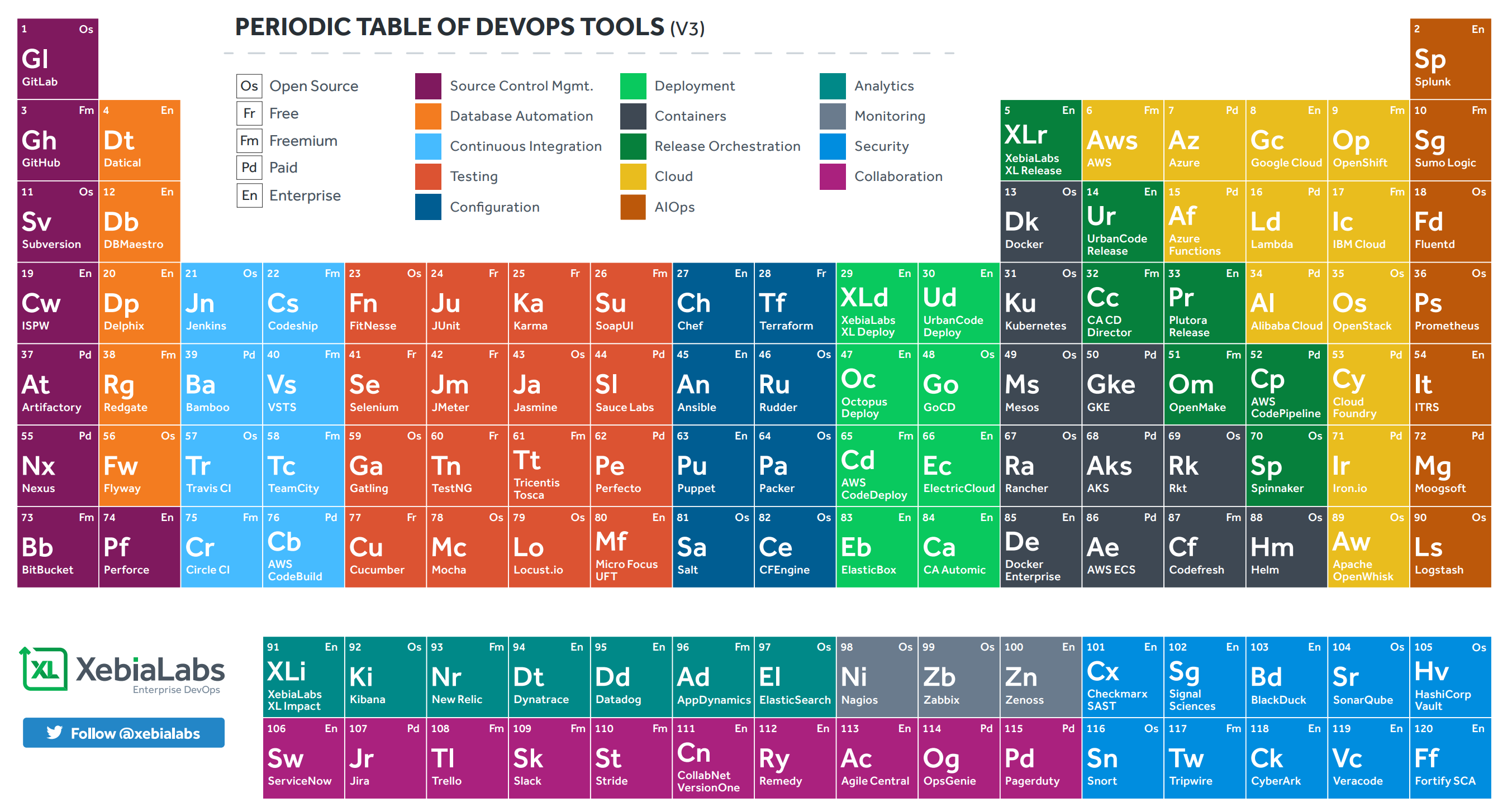}
\caption{The Periodic Table of DevOps-Tools}
\label{fig:periodic_table_of_devops-tools}
\end{figure}

\end{document}

%% file: body.tex
\section{Introduction}
\label{sec:introduction}
\begin{figure}[ht]
\centering
\includegraphics[width=0.9\textwidth]{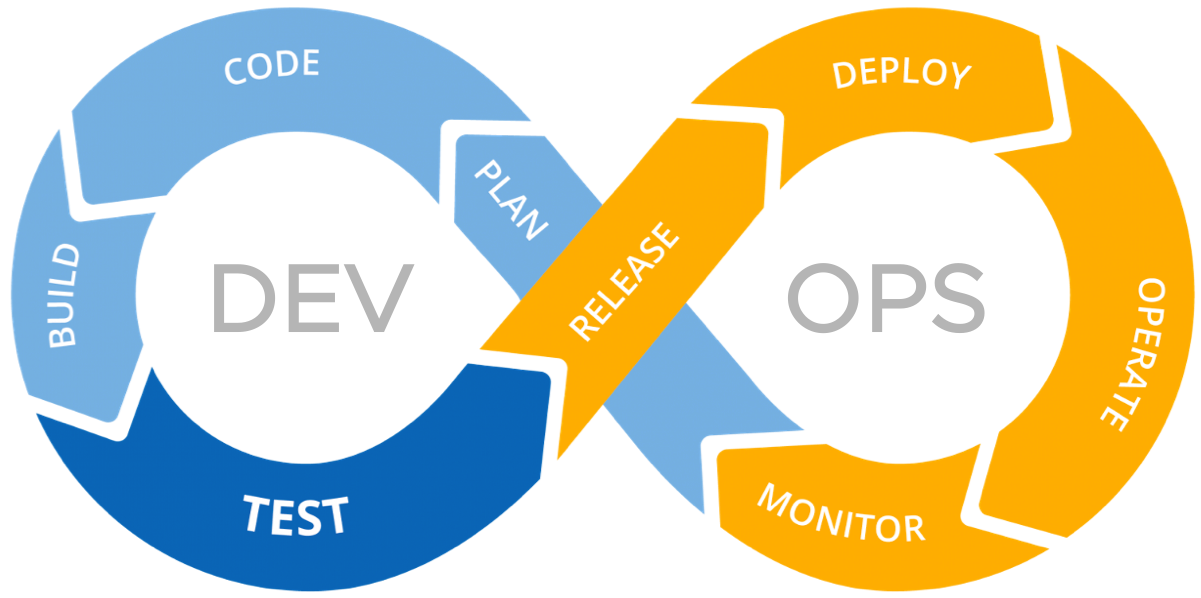}
\caption{The DevOps-Life-Cycle}
\label{fig:devsecops-life-cycle}
\end{figure}

Several studies,
e.g. from \textit{Gartner}~\cite{Gartner.2019,Gartner.2017},
continue to predict drastic numbers of \ac{iot} devices in use.
Some are dedicated to certain specialised environments,
such as industrial applications,
creating the \ac{iiot},
or automotive scenarios.
However,
these \ac{iot} and \ac{iiot} devices contain severe vulnerabilities,
often due to the fact that security solutions are not suited or not applied to \ac{iot} environments.
\textit{Mirai} and other botnets were capable of infecting up to \numprint{500000} devices with dictionary attacks and consequently use those devices to perform \ac{dos} attacks that heavily impacted the internet~\cite{Antonakakis.2017,Kolias.2017}.
In \ac{iiot} environments,
where devices are linked with \acp{cps},
attacks in the digital domain are capable of influencing the physical domain.
Thus,
vulnerabilities in the \ac{iot} can not only cause monetary loss,
but also physical harm to persons and assets.
At the same time,
the DevOps-method has been well-established in software life-cycle management.
A typical DevOps-life-cycle with the respective tasks mapped to the phases is shown in Figure~\ref{fig:devsecops-life-cycle}.
The DevOps methodology is well described and introduced in many companies.
There are multiple ways to implement the process depending on type of organisation and maturity of the organisation.
In general, three continuous phases can be distinguished in to which the eight steps of a DevOps-cycle are mapped as follows:
\begin{enumerate}
\item \ac{ci}: Code and build steps
\item \ac{cd}: Test, release, and deploy steps
\item \ac{cf}: Operate, monitor, and plan steps
\end{enumerate}
This terminology is generally accepted in industry.
DevOps as method is mainly used in controlled environments where connections are stable and infrastructure can be used by software to set up adhoc staging areas.
However,
most tooling specific for cloud or enterprise environments cannot be used for large deployed \ac{iot} systems.
Distributed deployment and feedback is difficult to achieve with the tools at hand.
Furthermore,
several studies show an unacceptable number of security related issues that pass undetected to deployment without the means for a quick recovery~\cite[page 32]{EBOS.2019}. \\ \par
This work presents a state of the of art regarding the area of DevOps,
with a focus on gaps in security as well as \ac{iot}-applicability in Section~\ref{sec:related_work}.
It is used as a framework for investigating these gaps and developing improvements to be implemented in the course of research project \ac{scratch} in Section~\ref{sec:solution_architectures}.

\section{Tooling for the DevOps-Cycle}
\label{sec:related_work}
DevOps is currently implemented by many companies, in different levels of maturity.
Proof of this is the abundance of available tooling and consultancy.
An overview by XebiaLabs~\cite{XebiaLabs.2020} illustrates this.
This overview of the different tools and the phase in which the are employed is shown in the periodic table of DevOps-tools,
provided by XebiaLabs see Figure~\ref{fig:periodic_table_of_devops-tools}~\cite{XebiaLabs.2020} in the appendix.
However,
tooling is not per definition applicable in the \ac{iot} domain and the main advantage of DevOps,
quickly implement updates and mitigate security breaches,
is lost.
The increasing issue of creating secure \ac{iot} systems puts an additional focus into the DevOps cycle.
There are tools readily available to build a DevOps Chain,
however,
specific \ac{iot} security objectives are not tackled by these tools.
DevOps does not provide sufficiently holistic solutions yet.
This section presents an overview of the three main phases of DevOps,
tools that can be used,
their shortcomings and recommendations to improve working methods and tools for a more secure DevOps life-cycle in the \ac{iot}-domain.

\subsection{Continuous Integration (CI)}
THe plan, code and build phases of DevOps are referred to as CI or continuous integration.
This is where the DevOps process initially starts and a product or piece of software is created. 
Scrum and other agile methods are used and teams do incremental code development (sprints) until a certain release or minimal viable product is created.
Code is regularly checked in to the main branch, called integration and builds of the code are made often on regular basis,
e.g daily.
Important aspects in this phase related to \ac{iot} and security are~\cite{franklin2020}:
\begin{enumerate}
\item Specific security related design constraints
\item System design or architecture
\item Security awareness agile team~\cite{Puppet.2019}
\item Code and vulnerability checking.
\end{enumerate}
Most companies are at this level and apply some type of continuous integration,
however,
not every organisation applies all of the 4 points mentioned.\\ \par
\textbf{{Recommendations:}}
The Design-Phase is a bit of a chaotic stage \cite{franklin2020} where agile working methods are common practice. Security by design is not a hollow phrase. It is important because most security features and future behaviour of a system is set in the design phase. Yet it is not common practice. A few simple methods and tools to improve the implementation of Security are:
\begin{enumerate}
\item Inject security design constraints as non functional requirements or stories into the process:
\begin{itemize}
\item In the design phase specific security features are not always injected by the customer or product owner, using the outcome of some good practice research e.g. from \ac{enisa}, will increase security awareness and contribute to a  more robust architecture. Some good practices are listed in the knowledge base of \ac{scratch}.
\end{itemize}
\item Design resilient architecture:
\begin{itemize}
\item If design constraints are reflected in the architecture or system design, it is likely that the system is better equipped in the operate phase to recover from security breaches.
\item \ac{iot} solution architectures can play an important role in keeping a system safe. If in the architecture methods are defined that increase the resilience of a system e.g. methods to securely interact with the system. 
\end{itemize}
\item Introduce a simple security analysis exercise for the team, e.g. using STRIDE:
\begin{itemize}
\item Number one on the list of practices that improve Security in the survey~\cite{Puppet.2019} is collaboration between security and development teams on threat models. A simple template developed in \ac{scratch} can be used for this purpose.
\end{itemize}
\item Pick an appropriate threat checking tool:
\begin{itemize}
\item Integrate security tools into the CI pipeline ensure that developers can be confident they’re not inadvertently introducing known, security problems into their code-bases \cite{Puppet.2019}.
It is recommended to extend threat checking through out the DevOps cycle,
e.g at release to check for changes in used libraries, at operate to check for new introduction of security issues.
\end{itemize}
\end{enumerate}

\subsection{Continuous Deployment (CD)}
For CD it is assumed that staging areas can be setup instantaneously,
the potential to be released software can undergo automatic testing in this staging area,
results of these ``final'' test lead to deployment or not.
It is clear that a gap exist here for \ac{iot} systems,
as it is impossible to setup adhoc \ac{iot} environments that mimic the real implementation~\cite{franklin2020}. \par
The deployment software stack itself is complex.
Common tools to operate software are containers (Docker), local virtual machines (VMware, VirtualBox) and cloud systems (AWS, Azure).
For orchestration, Vagarant and Kubernetes are well-established options.
The automated configuration is frequently realized with Ansibl, Puppet, Saltstack and Chef.\par
Transport (TLS) or application (HTTPS, SSH) layer security can be considered standard for communication in the deployment phase,
Section~\ref{ssec:if_for_end_node} discusses methods for pre- and late provisioning, thus confidentiality, integrity and authenticity are provided.
Several tools also include support for enterprise authentication, such as LDAP.
Security in the deployment phase is significantly based on (default) configuration, best practices and linting.
Section~\ref{ssec:if_for_end_node} discusses methods for pre- and late provisioning to cover the authenticity of a \ac{iot} device.
It is important to avoid insecure default configurations.
Vagrant, for example, used default credentials and a default key pair for SSH communication in the past.
It is also imperative to include linting tools to verify configurations during the deployment.
An example of such a tool is hadolint or kubesec.io.
Hadolint checks configuration files of docker containers for compliance with security best practices, whereas kubesec.io conducts an automated risk assessment of the configuration.
Furthermore, the choice and procurement of base systems may affect security. Several environments allow to share base systems among their users. Using base systems from untrusted sources imposes a security risk. Additionally, the integrity should be validated to ensure a secure basis for further deployment. An example for integrity protection is the container signing tool Portieris by IBM and Notary by Docker.
Compromised systems cannot be avoided completely by complying to best practices.
There have been, for example, 75 vulnerabilities be reported for Ansible between 2013 and 2020.
To increase resiliency defense-in-depth strategies may be applied and container firewalls as provided by vArmour may be set up.
Distributing firmware or configuration to a life \ac{iot} system ads to the complexity of these containers and the underlying design of the  \ac{iot} hardware, as a interrupted update can cause a loss of connection without immediate means to recover.\par
In the recent past, the severity of insecure deployment stacks was showcase in the Tesla Hack in 2018.
Tesla used a Kubernetes console that was not password protected.
Criminals found it and gained access to the orchestrated containers.
One of these containers included credentials for an AWS cluster, which was compromised subsequently.

\subsection{Continuous Feedback (CF)}
The operate, monitor, and plan-phase of the DevOps-cycle are referred to as \ac{cf}. In these phases, the system is operated in the intended fashion and its operation is monitored.

An \ac{iot} system in operation generates feedback (monitoring) information, some part of this feedback can lead to automatic adaptation or change in the \ac{iot} environment as a preventive act to mitigate an identified risk, Other feedback is fed back into development to improve on the system, and some feedback will be analysed by operation. The whole system of data collection and actions is a constant learning curve that if successfully implemented leads to a more stable system. \par

Among other \acp{kpi}, such as application performance and throughput, the operate and monitor-phases are analysing the systems for security-related incidents. In general, if the vulnerabilities have not been discovered in the code- or build-phase based on source code analysis or functional testing, they can be detected in the operate- and monitor-phases by either host- or network-based intrusion detection systems. A strong focus of security tools for DevOps lies on data management and the ensurance of privacy, e.g. by companies such as SignalSciences, HashiCorp, and CyberArk. CyberArk provides account management for privileged accounts aiming at the reduction of information leakage and misconfiguration. HashiCorp enables provisioning and securing cloud environments and SignalSciences provides \acp{waf}. Their focus lies on the ensurance of data security. However, as no complete security can be guaranteed for any system, it is desireable to detect attacks and intrusion if they occur. Snort performs network-based intrusion detection with a wide set of rules based on network packets. Tripwire provides a similar approach by creating snapshots of a given system's state and alerting the user in case of a change. Thus, intrusions and changes in behaviour can be detected.

There are, however, several disadvantages or drawbacks to such solutions. Since classic \acp{ids}, such as Snort and Tripwire, do not take into consideration the architecture of the systems, the rules for detecting intrusions are generic. Furthermore, the extent of an attackers influence, as well as their goals and methods, cannot be derived from such information. Getting insight about the intention and approach of an attacker would allow to better prepare for future attacks. Finally, rule-based \acp{ids} can only protect against attacks of which the signatures are known, meaning that unknown or stealthy attacks cannot be detected. \par

Within \ac{scratch} two tools are under development that improve Security in the Operate Phase,
Deception in IoT (Section~\ref{ssec:deception_for_iot}) and Anomaly Detection (Section~\ref{ssec:ad-based_ids})

\section{Solution Concepts}
\label{sec:solution_architectures}
This section presents individual solution concepts, addressing challenges in the DevOps-cycle brought up by security risks as well as the adaption of DevOps into \ac{iot}-environments.
A summary mapping the solutions to the respective phases is shown in Table~\ref{tab:solution_overview}.
\begin{table}[h!]
\renewcommand{\arraystretch}{1.3}
\caption{Mapping of Solutions to DevOps-phases}
\label{tab:solution_overview}
\centering
\scriptsize
\begin{tabular}{l c c c c}
\toprule
\textbf{Solution} & \phantom{a} & \textbf{CI} & \textbf{CD} & \textbf{CF} \\
 \cmidrule{3-5}
Pentesting IoT Devices (Section~\ref{ssec:active_security}) & & & \cmark & \\
Deception in \ac{iot} (Section~\ref{ssec:deception_for_iot}) & & & & \cmark \\
Anomaly Detection (Section~\ref{ssec:ad-based_ids}) & & & & \cmark \\
ID for the End Node (Section~\ref{ssec:if_for_end_node}) & & & \cmark & \\
Software / Firmware Updates (Section~\ref{ssec:sw_fw_updates}) & & & \cmark & \\
\bottomrule
\end{tabular}
\end{table}
Since project \ac{scratch} addresses singular challenges in \ac{iot}-based DevOps environments,
not every phase is covered by individual solutions.
Instead, concerns that were raised due to distributed DevOps applications as well as their implications on security,
are the main focus of \ac{scratch}.
The table shows that \ac{ci} is not addressed in the context of \ac{scratch} since most issues in terms of security integration and \ac{iot} applicability are in the \ac{cd} and \ac{cf} phases.
Generally,
consideration of security issues,
security by design,
could provide a more seamless integration and coverage of security objectives.
This would require systems and software to be developed in a fashion that integrates security integration as early as the design phase.
If tooling for security is applied in later stages,
the risks are realised in the \ac{cd} or \ac{cf} phase.

\subsection{Pentesting IoT Devices}
\label{ssec:active_security}
During a penetration test (or pentest for short), an application’s overall security posture is measured against simulated cyber attacks.
The result of a pentest are potential security vulnerabilities that could impact the application’s security posture in terms of confidentiality, integrity and availability.
Historically, pentests are performed before a new application or application’s feature is released so that there is less chance of potential vulnerabilities making it to production.
The tests performed during a pentest, are often performed manually by a pentester and with the support of tools. These tools can take many forms, ranging from tools that scan for security vulnerabilities fully automatic, to tools that allow manual interaction with an application component, in a way that it was not intended by the developers. Sometimes, and depending on the tooling available, a pentester might even write its own tools to allow testing of very specific cases.
Within the context of SecDevOps, performing a pentest before deployment would thus not fit the continuous deployment practice as a pentest is not fully automatic and would therefore not allow fully automatic deployments. 

Within the context of testing IoT devices, tools that automatically test security are scarce. This can be attributed to a couple of key properties that differentiate IoT devices from more traditional applications such as web or mobile applications: 

\begin{itemize}
    \item Diverse and embedded nature – IoT devices are often designed to have a single use (smart light bulbs, smoke detectors) and are often developed on hardware and software platforms tailored specifically for that use. As a result, there is not one but a highly diverse set of hardware and software platforms to take into account;
    \item Use of different physical layer communication technologies: IoT devices are connected to networks via a wide range of wireless links, such as Bluetooth Low Energy (BLE), 802.11, GSM/UMTS, LoRaWAN;
    \item Use of different application layer communication technologies: IoT devices tend to make use of machine to machine communication technologies such as MQTT and AMQP; 
    \item Large attack surface: IoT devices are part of a large and complex ecosystem. For example, a home alarm will be receiving input from motion detection, cloud and even mobile applications.
\end{itemize}

Due to the above-mentioned characteristics, fully relying on automated tools for IoT device testing is not yet feasible today. As such, validating security requirements through manual pentesting still remains a crucial part of securing an IoT device. There are a couple of challenges in pentesting of IoT devices. \par
First of all,
there is a lack of security verification requirements and testing guides for IoT devices.
For example, for pentesting traditional application such as web and mobile applications, OWASP provides security verification standards and testing guides that provide details on generic security verification requirements and how these can be validated through testing. For IoT devices, consolidated guides such as these do not yet exist.\par
Second, there is a lack of tooling to support pentesting activities of IoT devices.
For example, for testing the security of web applications, tools such as Burp Suite and OWASP Zap exist.
These tools provide features such as inspection and manipulation of HTTP request sent between client and server and automatic vulnerability scanning through fuzzing HTTP requests.
While these tools can also be used if the IoT device makes use of HTTP, unfortunately,
for many specific IoT technologies as mentioned above,
there is a lack of tooling. Interesting to note is that for many of the traditional pentest tools today more and more being automated as well.
For example, Burp Suite Enterprise Edition and OWASP ZAP’s Docker containers allow for an easy integration of these tool’s automatic vulnerability scanning features in a deployment pipeline.

\ac{scratch} aims to assist the community in establishing the foundations for validation security requirements of IoT devices through pentesting. In the short term, this will enable verifying the overall security posture of IoT devices, albeit through a laborious and mostly manual testing endeavor. However, by providing solid foundations for efficient pentesting \ac{scratch} hopes more and more tools will be able to provide automatic testing. \ac{scratch} aims to achieve this by focusing on the following three steps:
\begin{itemize}
    \item First, \ac{scratch} aim to create a repository of security verification requirements for IoT devices;
    \item Second, based on these requirements we aim to start with the creation of a testing guide that documents how these verification requirements be tested for specific technologies;
    \item Third, we invest in the creation of tooling that supports these testing activities.
\end{itemize}

\subsection{Deception in the IoT}
\label{ssec:deception_for_iot}
The earliest form of deception in the field of computers were honeypots, which were network port functioning as canaries to detect interaction. As no legitimate service was using this network port, every connection to the port would be illegitimate. Later the field of deception technology has been extended to different computer domains such as databases, memory stacks and network topology. A database could have canary tables or entries which raise alerts when accessed and computer memory could have canary entries on different stack locations, which, when being overwritten, indicate buffer overflow attacks. The network topology could change over time, utilizing Moving Target Defense (MTD), to confuse potential attackers and reduce the attack surface. In classical security operations an asymmetry between attackers and defenders exist, where defenders cannot afford to make any mistake, whereas the attacker has unlimited time and attempts to find a mistake. Deception may help to give the defender the advantage of detecting malicious attempts, while causing uncertainty and precaution for the attacker. Bell and Whaley coined a taxonomy for deception, where they distinguish two modes of deception, simulation or showing the false and dissimulation or hiding the truth. Further, according to this taxonomy, simulation can be described as either mimicking, inventing or decoying and dissimulation as either masking, repackaging or dazzling. In \ac{scratch}, deception strategies are researched for the IoT security domain. A large focus of IoT security lies on firmware security and update distribution, which could be improved using canaries for reverse engineering detection, canaries for memory corruption detection and feint patches in firmware updates to distract from the actual vulnerabilities the security update patches. Similarly, special pins on a hardware chip could be used as a tampering detection and disable the chip. How such deception and canary tokens could be planted into the firmware as part of the CI-pipeline is in scope of the \ac{scratch} research. Another focus of IoT security is the network security, where the application of the previously introduced MTD is being researched. One possible strategy could be that the addresses of each IoT device change in short time intervals.

\subsection{Anomaly-based Intrusion Detection}
\label{ssec:ad-based_ids}
Methods and tools for intrusion detection are well-established for end user devices. Such tools encompass anti-virus software and firewalls, furthermore network segmentation is a commonly applied method to secure systems. Generally, there are two methods to detect and prevent attacks: signature-based and anomaly-based. Signature-based intrusion detection is founded on the assumption that a given attack is characterised by a certain behaviour, e.g. a pattern of network packets. A signature-based \ac{ids} can scan for this pattern and consequently detect this kind of attack. Such systems are robust for known attacks, however, obfuscation techniques can make it hard for an \ac{ids} to detect the attack. Furthermore, novel attacks cannot be detected as no signature is known of them. Most common tools are signature-based.
\\ \par
Anomaly-based \ac{ids} learn models of a system's normal behaviour and discovery deviations from the normal behaviour. This allows them to detect formerly unknown attacks for which no signature is available. However, this commonly comes at the cost of a higher false positive rate, i.e. events that are incorrectly classified as an attack, compared to signature-based \ac{ids}. 
\\ \par
These approaches work well for known attacks and regular system behaviour. In the domain of the \ac{iot}, a heterogeneous landscape of devices, many of which work without user interaction, these preconditions are not met. Due to the variety of use cases, it is difficult to derive system models for anomaly-based \ac{ids}. Additionally, several devices contain several vulnerabilities so that signature management becomes difficult. Especially if attacks solely make use of valid commands and packets attacks are difficult to detect based on their signature. In order to tackle this issue, two dimensions are evaluated in an automated fashion: Timing and spatial distribution. Previous research has shown that in the \ac{iiot}, most behaviour is periodic or at least reoccurring~\cite{Lohfink.2020}. An algorithm called \textit{Matrix Profiles}~\cite{Yeh.2016a} can be used to extract sequences of characteristic behaviour, e.g. number of packets in and out, open connections, and detect anomalies. Since this algorithm is efficient in terms of computational time, this can be done for each device in a network, so that despite the heterogeneous systems, each device has an anomaly score to detect intrusions. The sequences are extracted by the \textit{Matrix Profile} algorithm in a sliding window fashion. Then a distance metric is calculated and the minimum of all distances for a given sequence is kept. A low minimal distance indicates the presence of a similar sequence in the time series, a high minimal distance indicates an outlier that can be an attack. Apart from regular behaviour in terms of timing, \ac{iot} networks often contain regular patterns of communication. These patterns can be extracted by considering the amount of in- and outbound connections, e.g. based on TCP sessions. If this value is taken in a periodic fashion, a time series is created that can be analysed for anomalies with the \textit{Matrix Profile} algorithm. \\ \par
As previous research shows, \textit{Matrix Profiles} are capable of perfectly, i.e. with neither false positives nor false negatives, detect attacks in novel industrial environments, such as the \ac{iiot}~\cite{Duque_Anton.2019a,Duque_Anton.2018}. Since \ac{iot} environments are similar in important characteristics, this approach in combination with the integration of relationships between devices is expected to detect attacks that are formerly unknown with a high accuracy. \\ \par
In addition,
machine learning methods have proven to successfully learn a normal model behaviour,
despite irregular human interaction,
and reliably detect attacks in network traffic data.
Especially \acp{svm} and Random Forests perform well with perfect or near perfect accuracy~\cite{Duque_Anton.2018b}.

\subsection{Secure deployment ID for IoT components}
\label{ssec:if_for_end_node}
Deploying new software to a IoT device has several challenges,
with secure identification being the first that needs to be addressed.  

\subsubsection{IoT identity provisioning}
It sounds simple, but is an essential question in the digital world: how do you prove ownership of a physical device? 
Consider you want to bring in some new IoT device into your home or office network.
Leaving aside the challenge of connecting it to WiFi, by assuming you just plugged it into your router, how do you connect to it?
And how do you make sure nobody else connects to it or,
even worse,
completely high-jacks the device?
Most of the times there is a centralised web platform where you register yourself and your devices.
But likely you are not the only user connecting a new device at the moment - how do you know which device belongs to which user?
There are a few ways to do this - and some compromise security more than others. 
Within \ac{scratch} we intend to look implementing a feasible solution for SMEs out of:
\begin{itemize}
    \item Late Stage Parameter Configuration~\cite{NXP_late}
    \item Pre-Provisioning Keys
\end{itemize}

There are large similarities within the usual identity processes. The device needs to hold a secret which is only known to the rightful owner.
When claiming ownership towards a management platform, both the user and the device are linked upon matching that secret.

\subsubsection{Late-stage Parameter Configuration}
If the device is as sophisticated as a laptop and provides some means of either input or output, the identity process can be achieved by multiple means like
\begin{enumerate}[(a)] 
\item entering a secret on the device which then is also entered on the management platform
\item displaying a random key generated on the device and send to the platform which is entered again on the management platform by the user
\end{enumerate}
If the device does not have such means of interaction 
\begin{enumerate}[(c)] 
\item using an additional device like a mobile phone which connects locally to the IoT device and acts as input/output provider is an option.
\end{enumerate}
As a variation of process the user could attach storage to the device holding such secret key. 
All these processes share a commonality - they can be performed post-distribution when the devices is within the target network. There are many other approaches and deviations to connect new devices, especially if already authenticated devices exist within the target network. They are not considered as they still cause the initial challenge of getting one device registered, which mostly happens by using one of the processes described above. 
The benefit of this process is to avoid the logistical challenges laid out in the \nameref{Pre-Provisioned Keys} section below.

\subsubsection{Pre-Provisioned Keys}
\label{Pre-Provisioned Keys}
Some IoT devices come with pre-provisioned key. The onboarding process is similar to a) - a secret already exists on the device (ideally some kind of secure element) and is provided either physically (e.g. printed on the box of the IoT device) or digitally (like USB sticks) to the user. While this process is quite intuitive and, depending on the target audience, simpler to perform the logistical process of matching the digital secret within the device and the externally available secret should not be underestimated.

\subsubsection{Identities can be stolen}
It is important to consider that identities can be stolen.
It happens to devices~\cite{DeviceId} as it happens to people~\cite{IdTheft}.
The same way you don't let your wallet with your credit card laying around in a cafe, you need to protect the device identity from being stolen. A simple text file on an SD card will likely not be enough. Considering you plan to use this to provide the next level of Netflix, Hulu or Amazon prime - you can be sure that this text file will be shared on the Internet quickly and people will abuse and  consume your services free of charge.
As a general rule its good to remember that the attack on the identity (and thus affiliated services) should be more expensive than purchasing the services legally. Secure elements are a proven solution here - at the costs of cents they protect already today high value assets from credit cards to passports and ID cards.

\subsubsection{Additional challenges to consider}
Having an architectural approach for your IoT identity provisioning solving the previously mentioned challenges is good. But usually it is only a part of the entire life cycle and environment to consider.

For the \textbf{life cycle} it must be assumed that devices are being de-provisioned (due to being broken, stolen, sold,  hacked, etc.). Some of those devices will be re-provisioned by a different user. Some devices need to be able to rollback to a trusted state and re-provisioned while considering that this can be an attack vector as well. The chosen identity solution needs to consider this. From a security point it must be possible to blacklist devices and not just trust anything.

Some \textbf{environmental challenges} come from the device and the enclosing network itself. Starting with the available bandwidth: Not every device is connected to broad band allowing Mega or even Gigabytes to be transferred. E.g. a Sigfox payload size is as little as 12 bytes~\cite{Sigfox.2017}. This imposes limits on ciphers to be used for secure transmission of the shared secrets.

Being \textbf{on- or off-grid} with regards to the power connection will impact the latency of communication (e.g. the device may not be always on due to power saving requirements), the computational power you can put into your provisioning may be limited due to the same reason as well.

The device may be behind \textbf{NAT or other firewall} setups not allowing direct communication as intended. A proxy may be required to ensure secure communication to the managing entity.

In general it is challenging to ensure not just secure identity provisioning, but maintaining the devices security through the entire life cycle. Another challenge the \ac{scratch} project is looking at, is to provide easy means of updating the device in the field. Re-reading the previously mentioned challenges may make you aware of the further challenges out there when managing IoT infrastructures. 

\subsubsection{Solution direction}
The solution direction \ac{scratch} aims at is to provide a easy and usable method for SMEs to get their IoT devices out to the customer without worrying about the logistical aspect pre-provisioned processes allowing a late-stage parameter configuration. For this the device will actively seek a connection to a pre-configured proxy server which then established a secured connection via a unique address. Once this connection is in place the centralized IoT server can negotiate the identity with the device and register it to the rightful owner. The credentials will be encrypted during transfer and at rest and where possible be stored in a secure element, given this is available.

\subsection{Software / Firmware updates for IoT }
\label{ssec:sw_fw_updates}
Keeping an \ac{iot} system safe throughout its life cycle needs some methods of interacting and updating the infrastructure~\cite{franklin2020,EBOS.2019}.
In the \ac{scratch} project we see secure identification as discussed in Section ~\ref{ssec:if_for_end_node} as a first step to solve. 
Apart from the ownership of the device, the device also
needs to know for sure that the firmware updates are
provided by the intended source, a requirement addressed by update authorization.
In this section this problem is addressed. 

Large Scale \ac{iot} deployment poses difficulties 
on software update security. Partially caused by the many different 
use cases, huge variety in actual devices and their capabilities 
and architectures. There are devices running advanced operating systems,
having \ac{mpus}, some run on powerful \ac{mcus} yet others have 
very restricted resources. And there are devices with combinations of
powerful \ac{mpus} and \ac{mcus}, each running their own firmware. 
This leads to many combinations where specific risks arise
or where requirements or objectives compete or even become
mutually exclusive~\cite{assured.2018}.

Platforms or tool kits can help manage these problems, but
might be limited in that they only cover some security aspects
or only applicable to a limited set of device types. 
It's pointed out by the \ac{enisa}~\cite{enisa} that several issues arise in
firmware/software updates of \ac{iot} devices, e.g.: Complex ecosystems,
fragmentation of standards and regulations and security integration.

A cryptography system which can be tailored to the 
specific use case, addresses these issues. 
Redwax ~\cite{redwax.eu} which provides a modular 
and decentralized approach, is demonstrated here with a 
very simple example. The principles of the Redwax architecture 
are appealing in our context because they aim for flexibility yet 
hide handles which require cryptography experts. 

In this scenario a new firmware is being sent to an \ac{iot} device and
the goal is to check the authenticity of the time stamp (which may include
a firmware sum) and whether the update is newer than the active firmware or not.
This happens on the \ac{iot}device. 
If the check fails the device can take appropriate actions, e.g. go to a specific
fail state, refuse the new firmware and keep the current etc. 

As an example we consider a Redwax time stamp server, 
which is located on premise where the firmware development takes place.
Access is under company control. 
The server is a hardware device equipped with a \ac{se}. 
The server is being setup  with timestamp front-end module, 
signing back-end module which uses a OpenSSL engine 
with the \ac{se} to provide signing with a private key. 
The server generates (root) certificates
which are to be installed on the \ac{iot} devices.

The time stamping is used to protect against the out-of-date
firmware problem,
where firmware could be provided with expiration date through
manifest,
and rollback attacks.
This is even more advantageous 
in a setting where regular software updates are part of a strategy. 
The secret key, being kept in the time stamp server, 
suffers minimal exposure and provides trust for the source. 
A next step is to extend this with \ac{pki} and implement it in e.g. 
an \ac{sme} setting. Emphasis is then on ``the integration and deployment
of their own cryptography system''\cite{redwax.eu}(architecture). 

In a next phase research will be done on how to cope with the specific 
\ac{iot} issues like non continuous connections, 
low bandwidth and hardware limitations. 
It is not given that we will reach the end goal of secure deployment of 
software containers and at the same time having a fail safe mode implemented 
in all \ac{iot} devices. However, the goal is clear and for any gap identified 
alternative solutions might provide a mitigation.

\section{Conclusion}
\label{sec:conclusion}
The DevOps-cycle has been successfully established for the application in IT- and cloud-based environments. Such environments are easy to manage in a centralised fashion, creation, roll-out, and management of code can be performed with established tools. However, the de-centralised nature of \ac{iot} environments makes it difficult to apply the standard tools presented in this work. Furthermore, it opens up security issues, in the roll-out of identities and software, but also in monitoring the networks for intrusions. Such gaps in available tooling are addressed by the research project \ac{scratch}, which focuses on selected demands not yet met by standard tooling. Such tools include connecting to edge nodes and uploading identities to them,securely updating software to non-classic IT-devices as well as intrusion detection methods for novel threats and threat intelligence. In a next step, these solution approaches are integrated into test case environments and evaluated with realistic scenarios.

\section*{Acknowledgements}
This work has been supported by ITEA3 through project SCRATCh (label 17005) with funding from:
\begin{itemize}
    \item The Federal Ministry of Education and Research (BMBF) of the Federal Republic of Germany, within the project SCRATCh (01IS18062E, 01IS18062C). 
    \item Netherlands Enterprise Agency
    \item The regional institute for research and innovation of Brussels Belgium, Innoviris.
\end{itemize}
The authors alone are responsible for the content of the paper.

%% file: main.bbl
\begin{thebibliography}{10}
\bibitem{DeviceId}
Bradford, V.: Why device id may not be enough to stop fraud
\end{thebibliography}
